\documentclass[12pt,preprint,a4paper]{aastex}
\usepackage{graphics,epsf}
\usepackage{amsmath}                
\usepackage{amsfonts}               
\usepackage{amssymb}                
\usepackage{epsfig}                 
\usepackage{subfigure}
\usepackage{float}
\usepackage{subfigure}

\newcommand{\cm}{{~\rm cm}}
\newcommand{\s}{{~\rm s}}
\newcommand{\km}{{~\rm km}}
\newcommand{\g}{{~\rm g}}
\newcommand{\K}{{~\rm K}}
\newcommand{\erg}{{~\rm erg}}
\newcommand{\yr}{{~\rm yr}}

\begin{document}

\title{Wave-driven stellar expansion and binary interaction in pre-supernova outbursts}

\author{Liron Mcley\altaffilmark{1} and Noam Soker\altaffilmark{1}}

 \altaffiltext{1}{Department of Physics, Technion -- Israel Institute of
Technology, Haifa 32000 Israel; lironmc@tx.technion.ac.il, soker@physics.technion.ac.il.}

\begin{abstract}
We suggest that the main outcome of energy leakage carried by waves from the core to the envelope of pre-collapse massive stars is
envelope expansion rather than major mass ejection. We show that the propagating waves add to the pressure in the envelope
at radii smaller than the radius where the convection driven by the waves becomes supersonic, the driven radius $r_{\rm d}$.
The extra wave pressure and wave energy dissipation lead to envelope expansion.
Using the numerical code MESA we show that the envelope expansion absorbs most of the energy carried by the waves.
A possible conclusion from our results is that pre-explosion outbursts (PEOs) result from a binary companion accreting mass
from the extended envelope and releasing a huge amount of energy.
The accreting companion is likely to expel mass in a bipolar morphology, e.g., jets and an equatorial ring.
\end{abstract}


\section{INTRODUCTION}
\label{sec:introduction}

Vigorous nuclear burning in the core of evolved stars can leave almost no signs on their surface.
The huge amount of nuclear energy is channelled rather to gravitational energy of the core, such as in core helium flashes
at the termination of the red giant branch (RGB) and shell helium flashes on the asymptotic giant branch (AGB),
or to neutrino lose in late stages of massive stars.
A small fraction of the nuclear energy, if it reaches the surface, can cause substantial envelope disruption.
The fraction of nuclear energy reaching the surface and its influence on the envelope are open questions.

Consider convection that is triggered by core helium flash at the tip of the RGB.
It is thought that the convection reaches up to the Hydrogen shell that provides an entropy barrier against mixing;
the mixing can trigger hydrogen burning (\citealt{Campbell2010} and references therein).
However, off-center ignition of helium because of neutrino cooling prior to ignition (e.g., \citealt{Mocak2009}),
or a weaker barrier in low metallicity stars (\citealt{Campbell2008}; \citealt{Suda2010} for $[Fe/H] < -2.5$),
lead to some mixing.
As a result of this mixing ignition of a large amount of hydrogen occurs in these RGB stars
(\citealt{Mocak2008}; \citealt{Mocak2009}; \citealt{Mocak2010}).
\cite{Bearetal2011} deposited the energy at the bottom of the envelope, and showed that suffice that
few percents of the energy released by the hydrogen burning leak to the envelope to cause a substantial envelope expansion.
A large stellar expansion is found also in some calculations of shell helium flashes (thermal pulses) along the AGB
(e.g., \citealt{Schlattl2001}; \citealt{Boothroyd1988}).

Vigorous core nuclear burning of neon, oxygen, and silicon occurs years to hours prior to core collapse supernova (CCSN) explosions,
and triggers convection in the core.
\cite{QuataertShiode2012} and \cite{ShiodeQuataert2014} proposed that waves excited by core convection carry a small fraction of
the nuclear energy released in the vigourous nuclear burning, and deposite most of it in the outer layers of the envelope.
\cite{Soker1992} considered the propagation of p-waves through the convective envelope of an AGB star and noticed that their amplitude grow by a large factor
when reaching the outer regions of the envelope. \cite{Soker1992} further argued that the high amplitude of the p-waves can substantially increase the mass loss rate.
In that study the p-waves where excited by a massive planet or a brown dwarf orbiting deep in the AGB envelope.
The wave luminosity was low, and convective viscosity did not influence much the wave propagation \citep{Soker1993}.

In the study of \cite{QuataertShiode2012} the p-wave luminosity can be super-Eddington, and they argue for p-waves
dissipation mainly in the outer layers of the envelope. The energy dissipation in their model leads to ejection of the outer layers of the envelope, or to expansion
in case of a WR star \citep{ShiodeQuataert2014}.
Their mechanism belongs to a group of single-star mechanisms (e.g., \citealt{Shaviv2000,Shaviv2001,Owockietal2004}),
to explain super-Eddington outbursts from stars, or increased mass loss rate due to mass loss by neutrinos \citep{Moriya2014}.

\cite{Soker2013} on the other hand, with the motivation to explain the pre-explosion outburst (PEO) of SN~2010mc \citep{Ofeketal2013b},
suggested that the p-waves lead to a large and rapid envelope expansion rather than a large mass loss episode.
This expansion triggers mass transfer onto a secondary star postulated to orbit the progenitor of SN~2012mc.
Most of the extra energy of the PEO in SN~2010MC, according to the binary model \citep{Soker2013}, comes from the accretion of
$M_{\rm acc} \simeq 0.1 M_\odot$ onto the secondary star.
As well, the gas outflowing at $v_{\rm ej} \sim 2000 \km \s^{-1}$ was launched from the accreting secondary star, most likely
in a bipolar outflow.

In this paper we study in greater detail the influence of energetic p-waves propagating through the envelope of a massive evolved star.
The stellar models to be used in the paper are described in section \ref{sec:stellar}.
In section \ref{sec:pwaveP} we estimate the envelope response by considering the energy density of the waves, and in section \ref{subsec:Meject}
we consider the evolution of wave energy dissipation with time.
In section \ref{sec:numerical} we insert energy into stellar models and follow their evolution thereafter.
We are interested in cases where according to \cite{ShiodeQuataert2014} substantial mass is ejected.
In most of their WR models no mass is ejected during oxygen burning, and in cases where mass is ejected it amounts to only $<0.01 M_\odot$.
Hence we do not consider WR models.
Our summary is in section \ref{sec:summary}.

\section{STELLAR MODELS}
\label{sec:stellar}

\begin{figure}[ht]
\centering
       \includegraphics[width=0.45\textwidth]{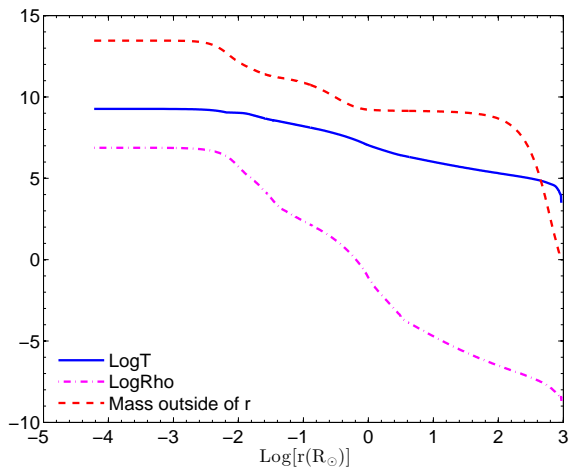}
\caption{The density (in $g \cm^{-3}$, logarithmic scale), temperature (in K, logarithmic scale), and mass (in solar mass) outside a radius $r$, as function of radial coordinate
in a red supergiant (RSG) star at the beginning of its core oxygen burning.
It has a present mass of $13.46 M_\odot$, a radius of $932 R_\odot$,
and effective temperature of $T_{\rm eff}=3145 \K$.
The envelope from $4 R_\odot$ outward is convective.
Its initial mass was $M_0=15M_\odot$ and it had a solar initial composition ($Z=0.02$).
}
\label{figure:RSGModel}
\end{figure}

\begin{figure}[ht]
\centering
\includegraphics[width=0.45\textwidth]{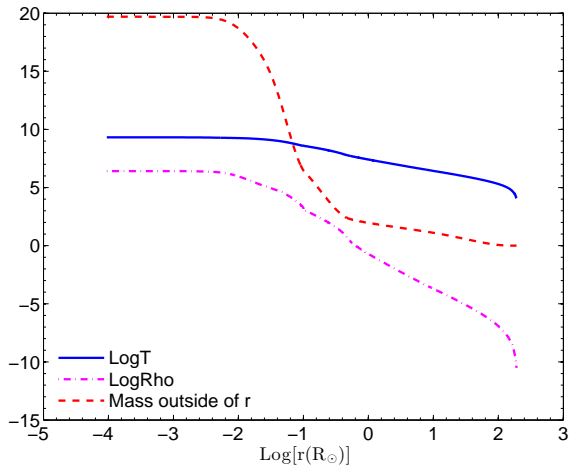}
\caption{The density (in $g \cm^{-3}$, logarithmic scale), temperature (in K, logarithmic scale), and mass (in solar mass) outside a radius $r$, as function of radial coordinate in a blue supergiant (RSG) star
at the beginning of its core oxygen burning. It has an initial mass of $M_0=40M_\odot$ and initial low metallicity of $Z=0.002$.
The stellar mass is $19.7 M_\odot$, a radius of $191 R_\odot$, and effective temperature of $T_{\rm eff}=1.20 \times 10^4 \K$.
The envelope from $20 R_\odot$ outward is convective.
}
\label{figure:BSGModel}
\end{figure}

We evolve two stellar models with the Modules for Experiments in Stellar Astrophysics (MESA; \citealt{Paxton2011}) version 5819.
The mixing length parameter is $\alpha=1.5$, Ledoux criterion for convection is used, and the so-called 'Dutch' scheme (e.g., \citealt{Nugis2000, Vink2001}) for mass loss is used
(based on \citealt{Glebbeeketal2009}).
The models at the beginning of their respective core oxygen burning are shown in Figs. \ref{figure:RSGModel} and \ref{figure:BSGModel}.

  The red supergiant (RSG) model has a ZAMS mass of $15M_{\odot}$ and a solar metallicity ($Z=0.02$) at zero-age main sequence (ZAMS).
We present it at its O-burning phase at an age of $1.2\times10^7\yr$ where its mass is $13.46M_{\odot}$, its radius is
$R_{\rm *}=932R_{\odot}$, its luminosity is $L_{\rm *}= 7.6\times10^4 L_{\odot}$ and its effective temperature is $T_{\rm eff}=3.1\times10^3 \K$.
The blue supergiant (BSG) model has a ZAMS mass of $40 M_{\odot}$ and metallicity of $z=0.002$ as used by \cite{ShiodeQuataert2014}
for a similar model.
We present it also at its O-burning phase at an age of $5.2\times10^6\yr$ where its mass just before explosion is $19.7M_{\odot}$, its radius is
$R_{\rm *}=191 R_{\odot}$, its luminosity is $L_{\rm *}= 6.7\times10^5 L_{\odot}$ and its effective temperature is
$T_{\rm eff}=1.2\times10^4 \K$.

\section{THE IMPORTANCE OF P-WAVE PRESSURE}
\label{sec:pwaveP}

\subsection{Energy content}
\label{subsec:Econtent}

\cite{Soker2013} calculated the ratio of wave to thermal energy.
The waves carry energy with a luminosity of $L_{\rm wave} \simeq 4 \pi r^2 e_w c_S$, where $e_w$ is the
energy density of the waves and $c_S$ is the sound speed.
The local thermal energy density of the gas is $e_{\rm th} \simeq  \rho c_S^2$, such that
\begin{equation}
    \frac {e_w}{e_{\rm th}} \simeq
    \frac{L_{\rm wave}}{L_{\rm max,conv}},
\label{eq:enden1}
\end{equation}
where
\begin{equation}
L_{\rm max,conv} = 4 \pi \rho r^2 c_s^3,
\label{eq:lmaxconv1}
\end{equation}
is the maximum power that subsonic convection can carry as used by \cite{QuataertShiode2012}.
The radius $r_{d}$ (marked $r_{\rm ss}$ by  \citealt{ShiodeQuataert2014}) is where the convection driven by wave energy deposition becomes supersonic and likely initiates
an outflow according to \cite{QuataertShiode2012}.
It is given by $L_{\rm wave} = L_{\rm max,conv} (r_{d})$.
\cite{QuataertShiode2012} give the values of $L_{\rm wave}$ and $L_{\rm max,conv}$ for their model, from which \cite{Soker2013} deduced that
${e_w}/{e_{\rm th}}=0.1$,  $0.3$, and $1$ at $r/R_\ast =0.07$, $0.1$, and $0.3$, respectively,
where $R_\ast=1700 R_\odot$ is the stellar radius of their model.
This implies that inner layers of the envelope at $r \simeq 0.1 R_\ast$ are substantially influenced by the waves, and will expand and absorb energy.
As the expansion time scale is similar to the waves-propagation time, \cite{Soker2013} argued that the envelope has time to respond and absorb most of the
energy carried by the waves, and that strong shocks will not be formed by the waves.

\begin{figure}[ht]
\centering
       \includegraphics[width=0.45\textwidth]{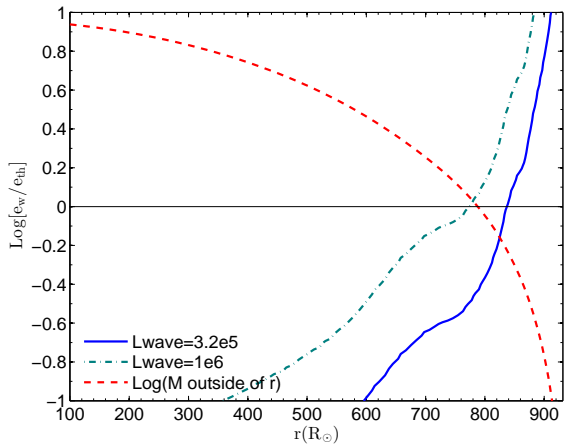}
\caption{The ratio of wave to thermal energy density (eq. \ref{eq:enden1}), for a wave luminosity of
$L_{\rm wave}=3.2 \times 10^5 L_\sun$ (solid line) and $L_{\rm wave}= 3 \times 10^6 L_\sun$ (dash-dotted line),
inside the RSG model presented in Fig. \ref{figure:RSGModel}.
The dashed line shows the mass outside of radius $r$ in logarithmic scale and solar units.
}
\label{figure:RSGWave}
\end{figure}

We here present the ratio of wave energy density, $e_{\rm {w}}$, to thermal energy density, $e_{\rm {th}}$, in the two stellar models described in section \ref{sec:stellar},
and at the beginning of their respective oxygen burning phase.
In Fig. \ref{figure:RSGWave} we present the ratio of wave to thermal energy density according to equation
(\ref{eq:enden1}) for two wave powers for the RSG stellar model at the beginning of its core oxygen burning (model in Fig. \ref{figure:RSGModel}). From table 2 of
\cite{ShiodeQuataert2014} for the non-rotating model we find the wave power
of the RSG model to be $L_{\rm wave,n}=3.2 \times 10^5 L_\odot$. We also show the ratio for a more powerful wave with $L_{\rm wave,n}=1 \times 10^6 L_\odot$.
According to \cite{QuataertShiode2012} driven outflow takes place from the radius where the ratio is 1 and outwards.
For the $L_{\rm wave}= 3.2 \times 10^5$ case (solid line) this occurs at
$r_d=837 R_\sun \simeq 0.90 R_\ast$, where $R_\ast$ is the stellar radius.
We note that already at $r=770 R_\sun$ the energy density ratio is ${e_w}/{e_{\rm th}} \simeq 0.3$,
and this ratio becomes ${e_w}/{e_{\rm th}} \simeq 0.5$ at $r=810 R_\sun$.
The waves energy density causes pressure enhancement that pushes the envelope outwards on the same time scale that the waves propagate.
This is the base for our claim that waves will cause envelope expansion rather than massive gas ejection.

\begin{figure}[ht]
\centering
\includegraphics[width=0.5\textwidth]{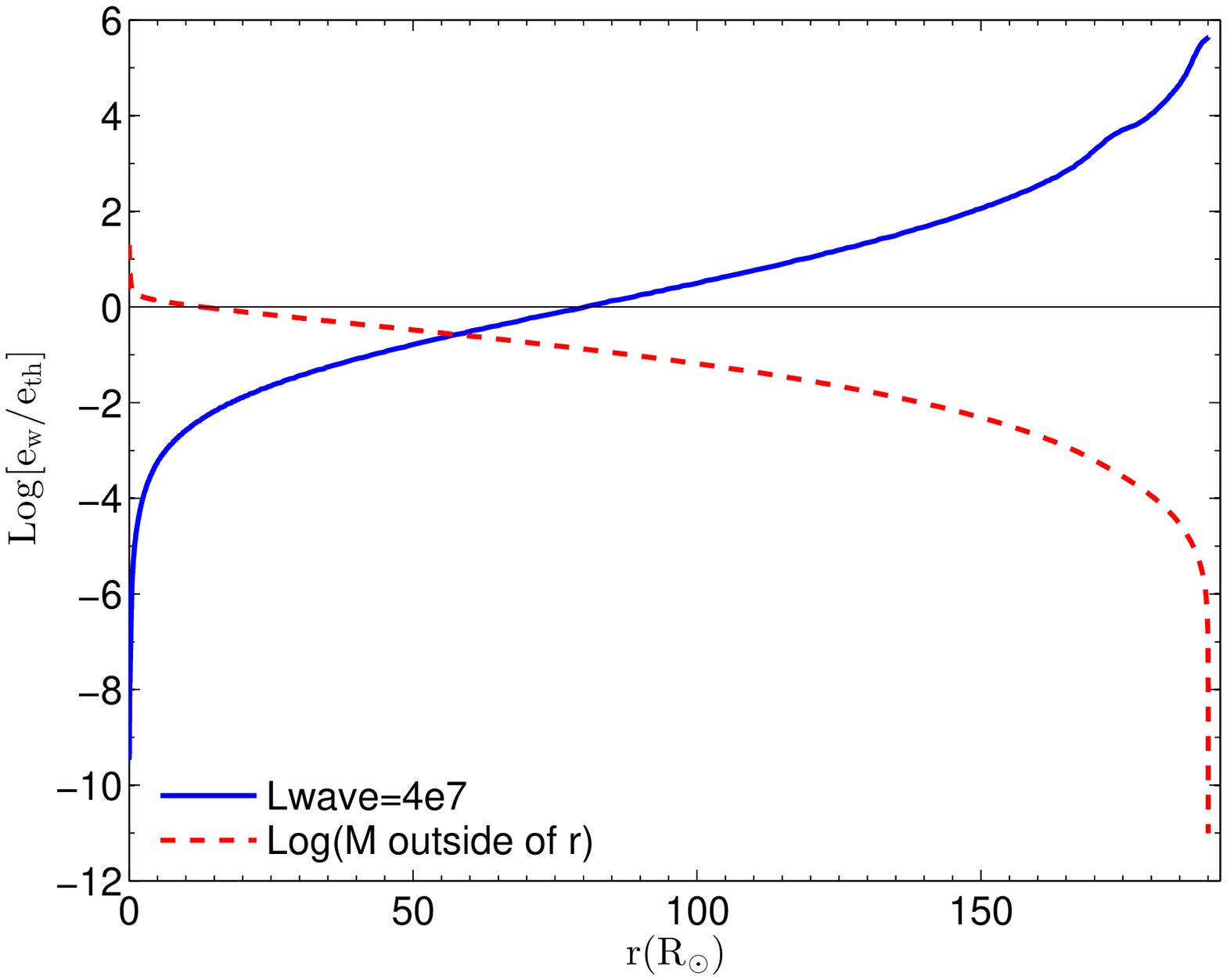}
\caption{Like Fig. \ref{figure:RSGWave} but for our blue supergiant (BSG) stellar model as presented in
Fig. \ref{figure:RSGModel}, and a different value of the waves power of $L_{\rm wave}= 4 \times 10^7 L_\sun$.
}
\label{figure:BSGWave}
\end{figure}

In Fig. \ref{figure:BSGWave} we present the energy density ratio for a blue supergiant (BSG) stellar model (model in Fig. \ref{figure:BSGModel}).
The waves power is according to table 2 of \cite{ShiodeQuataert2014}, $L_{\rm wave} =4 \times 10^7 L_\sun$,
for which the energy density ratio equals 1 at $r_d=80 R_\odot$.
However, below this region the pressure of the wave is already significant.
The wave energy density is third that of the thermal pressure already at $r\simeq 60 R_\sun$.
Such an increase in pressure will cause the envelope to expand on a dynamical time.

The RSG and BSG models presented above show that already below $r_{d}$ the waves will substantially increase the pressure,
and hence lead to the expansion of the envelope.
This expansion absorbs energy on the expense of the wave power. As the expansion starts deeper than $r_{d}$, the binding energy of the envelope there is larger both
due to high mass and to smaller radius. This implies that a small expansion of deeper layers can absorb as much energy as the ejection of the outer layers of the envelope can.
Based on these estimates we argue that most of the wave energy goes to envelope expansion.

\subsection{Pressure gradient}
\label{subsec:Egrad}

The pressure due to the propagating sound waves results from the oscillatory motion of the gas and amounts to $P_w \simeq e_w$.
As the waves propagate in the radial direction, they exert force along the radial direction proportional to the pressure gradient
\begin{equation}
\frac{dP_w}{dr} \simeq  \frac {d}{dr} {e_w} =  \frac {d}{dr} \left( \frac {L_{\rm wave}}{4 \pi r^2 c_s } \right)
={e_w}  \frac {d}{dr} \ln \left( r^2 c_s \right)^{-1} + {e_w}  \frac {d}{dr}{\ln L_{\rm wave}}.
\label{eq:gradpw1}
\end{equation}
The profiles in the envelops of red giant stars (RGB, AGB, RSG) can be approximated as
\begin{equation}
\rho \propto r^{-\xi}, \qquad P \propto r^{-(\xi+\beta)}, \qquad T \propto r^{-\beta}, \qquad {\rm for} \quad \xi \simeq 2-2.5 , \quad \beta \simeq 1.
\label{eq:profile}
\end{equation}
Using these relations and assuming constant wave luminosity we derive the ratio of the wave to thermal pressure $P_{\rm th}$ gradients
\begin{equation}
\frac{dP_w/ dr}{d P_{\rm th} /dr} \simeq  \frac{e_w}{P_{\rm th}} \frac{2-0.5 \beta}{\xi + \beta} \simeq 0.5 \frac{e_w}{e_{\rm th}} \qquad {\rm for} \qquad \frac{dL_{\rm wave}}{dt}=0
\label{eq:gradpwpt}
\end{equation}

The high luminosity wave phase starts in a relatively short time compared to the dynamical time of the outer envelope.
As the waves increase their luminosity at the beginning of each nuclear burning phase we have $ dL_{\rm wave} / dt > 0$.
We take the increase to maximum luminosity time $t_r$, to last a time somewhat shorter than the corresponding nuclear burning phase
$t_r \la t_{\rm fusion}$.
The last term in equation (\ref{eq:gradpw1}) reads
\begin{equation}
{e_w}  \frac {d}{dr}{\ln L_{\rm wave}} = {e_w} \frac {dt}{dr} \frac {d}{dt}{\ln L_{\rm wave}}   \simeq
 {e_w}\frac{1}{c_S t_r} \simeq  \frac {e_w}{r} \frac {t_{\rm dyn}(r)} {t_r},
\label{eq:dldt1}
\end{equation}
where $t_{\rm dyn}(r) \simeq r/c_S(r)$ is the dynamical time at radius $r$. We find
\begin{equation}
\frac{dP_w/ dr}{d P_{\rm th} /dr} \simeq  \frac{e_w}{P_{\rm th}} \frac{2-0.5 \beta + t_{\rm dyn}(r)/t_r}{\xi + \beta} \simeq  \frac{e_w}{e_{\rm th}} \qquad {\rm for} \qquad \frac{dL_{\rm wave}}{dt} = \frac {L_{\rm wave}}{t_r}.
\label{eq:gradpwptL}
\end{equation}
For a massive star $M\ga 10 M_\odot$ at a radius of $\sim 1 AU$ the dynamical time is $t_{\rm dyn} \sim$weeks$-$months, about equal to $t_r$. We used $t_{\rm dyn} \sim t_r$ in the last equality in equation (\ref{eq:gradpwptL}).
This shows that even without dissipation, the wave can cause the envelope to expand and absorbs energy from the propagating waves.
Part of the wave energy already dissipates to thermal energy at the damping radius $r_{\rm dump}<r_d$ \citep{ShiodeQuataert2014}, a process that increase thermal pressure
directly. As a result of these processes envelope expansion will take place already below $r_d$.

\section{ENERGY CONSIDERATION}
\label{subsec:Meject}

In the previous two subsections we argued that the pressure formed by the gas oscillatory motion set by the sound waves is sufficient to cause substantial envelope expansion,
before even wave dissipation is considered.
Here we argue that even if this pressure is neglected and only wave dissipation is considered, envelope expansion will be the main effect, rather than major mass ejection.
Of course, mass loss rate will increase as a result of envelope expansion and higher luminosity (e.g. \citealt{MauronJosselin2011}),
but the main effect is envelope expansion.

\cite{ShiodeQuataert2014} take the mass outflow rate from the radius $r_{d}$ where the waves drive the outflow to be
\begin{equation}
\dot M_d = 4 \pi r_{d}^2 \rho(r_{d}) c_S(r_{d}).
\label{eq:md1}
\end{equation}
They take the wind to be blown with the escape velocity from the star  $v_{\rm esc}$. Since $v_{\rm esc}> c_S(r_{d})$,
they conclude that the mass outflow rate from the star must be $\dot M_{\rm esc} < \dot M_d$.
They dismiss the possibility of envelope expansion as suggested by \cite{Soker2013} because
the total wave energy deposited at $r_{d}$ generally exceeds the binding energy of the star at $r>r_{d}$.
We bring here a counter argument.

We can consider two limiting cases in the wave-dissipation mechanism. We define $t_d$, the wave expansion time from $r_{d}$ to the surface
\begin{equation}
t_d \simeq \frac {R_\ast - r_{d}}{c_S(r_{d})} \sim \frac{R_\ast}{v_{\rm esc}} \simeq t_{\rm dyn}({R_\ast}),
\label{eq:td1}
\end{equation}
where the second equality is an order of magnitude one.
We also take $t_{\rm fusion}$ to be the duration of the wave excitation phase.
The ratio of wave energy $E_{\rm wave}=L_{\rm wave}  t_{\rm fusion}$ to binding energy  of mass residing at $r> r_{d}$ is
\begin{equation}
\frac{E_{\rm wave}}{E_d} \simeq \frac {L_{\rm wave}  t_{\rm fusion}} {4 \pi r_{d}^2 \rho(r_{d}) c_S^2(r_{d}) (R_\ast - r_{d})}
= \frac {t_{\rm fusion}} {(R_\ast - r_{d})/c_S(r_{d})} \simeq \frac{t_{\rm fusion}}{t_d},
\label{eq:Ed1}
\end{equation}
where we used the definition of $L_{\rm max,conv} (r_{d})=L_{\rm wave}$ from equation (\ref{eq:lmaxconv1}).

The two limiting cases are $t_{\rm fusion} \ll t_d$ and $t_{\rm fusion} \gg t_d$.
When $t_{\rm fusion} \ll t_d$, as is the case for exploding giants and the oxygen burning phase that lasts for few months, we have $E_{\rm wave} < {E_d}$.
The wave will not eject mass, but rather inflate the outer envelope.
Since the dynamical time of the star is longer than the time to explosion, the star will not have time to response,
and no pre-explosion outburst (PEO) will take place in this case.

When $t_{\rm fusion} \gg t_d$ then $E_{\rm wave} \gg {E_d}$.
However, the radius$r_{d}$ moves inward in mass coordinate, and we need to consider the integrated mass residing above $r_{d}$.
Consider the case $t_{\rm fusion} \gg t_d$. After a time $t_d$ the waves reach the stellar surface and deposited energy
of $E = L_{\rm wave}  t_d \simeq E_d$ in the layer residing above $r_{d}$. This energy, by definition, is sufficient to cause a major expansion
of this layer. And since $t_d$ is about the dynamical time, the layer has the time to response and expand.
Therefore, both the density and the temperature, hence the sound speed, drop at the initial radius $r_{d}$.
By the definition of $r_{d}$ (below eq. \ref{eq:lmaxconv1}), the new radius $r_{d}$ must move inward in mass coordinate to maintain the equality
$L_{\rm max,conv}=L_{\rm wave}$.
Our conclusion is that once the gas reaches the outward speed $\sim c_S (r_{d}) < v_{\rm esc}$, it will absorb no more energy as wave energy is deposited in further inner envelope layers.
The gas cannot escape. The star suffers a major expansion, but no massive ejection take place.

We can summarize this section by strengthening the argument of \cite{Soker2013} that high power waves excited near the core
will cause major stellar expansion rather than a major mass ejection.
Both the pressure caused by the gas oscillatory motion below $r_{d}$ and wave dissipation in the outer envelope contribute to this expansion.
Of course, when the wave power is set to values much larger than those given in table 2 of \cite{ShiodeQuataert2014}, some mass will eventually be ejected.
We show this in the next section.

\section{NUMERICAL DEMONSTRATIONS}
\label{sec:numerical}
\subsection{Numerical procedure}
\label{subsec:procedure}
We use the two models described in section \ref{sec:stellar}, and shown in Figs. \ref{figure:RSGModel} and \ref{figure:BSGModel} at the beginning of core oxygen burning.
When we reach the stage of core oxygen burning we inject wave energy according to the following prescription.
(1) The wave luminosity, duration, and total energy, for the numerical runs, $L_{\rm wave,n}$, $t_{\rm wave,n}$, and $E_{\rm wave,n}=L_{\rm wave,n}t_{\rm wave,n}$,
respectively, are taken from table 2 of \cite{ShiodeQuataert2014}.
(2) We look for the driven radius $r_{d}(0)$ where at $t=0$ the wave luminosity equals the maximum energy that can be carried by convection
$L_{\rm max,conv} (r_{d}) =L_{\rm wave,n}$. We start to inject energy into the stellar model shell at that radius with the power of $L_{\rm wave,n}$.
(3) The star expands and the new radius $r_{d}(t)$ where $L_{\rm max,conv} (r_{d}) =L_{\rm wave,n}$ moves inward in mass and radial coordinates.
We inject the energy in the numerical-shells residing in the range $r_{d}(t)$ to $r_{d}(0)$.
(4) We continue with energy injection until $t=t_{\rm wave,n}$. We then follow the star evolution toward explosion.

This procedure is not fully consistent, as MESA in the mode we use it cannot handle the ejection of mass. However, we only want to show that
the star can arrange itself into a new structure with the injected energy.
In the region where the energy is injected in our models the dynamical time is shorter than the wave energy injection phase,
and the stellar structure in this region and inward is treated consistently.
The outer regions, where the extended envelope is formed, is likely to be dynamically expanding at slow velocity
(less than escape velocity), rather than be in hydrostatic equilibrium.

\subsection{A $15 M_\odot$ red supergiant model}
\label{subsec:RSG}

\begin{figure}[ht]
\centering
   \includegraphics[width=0.65\textwidth]{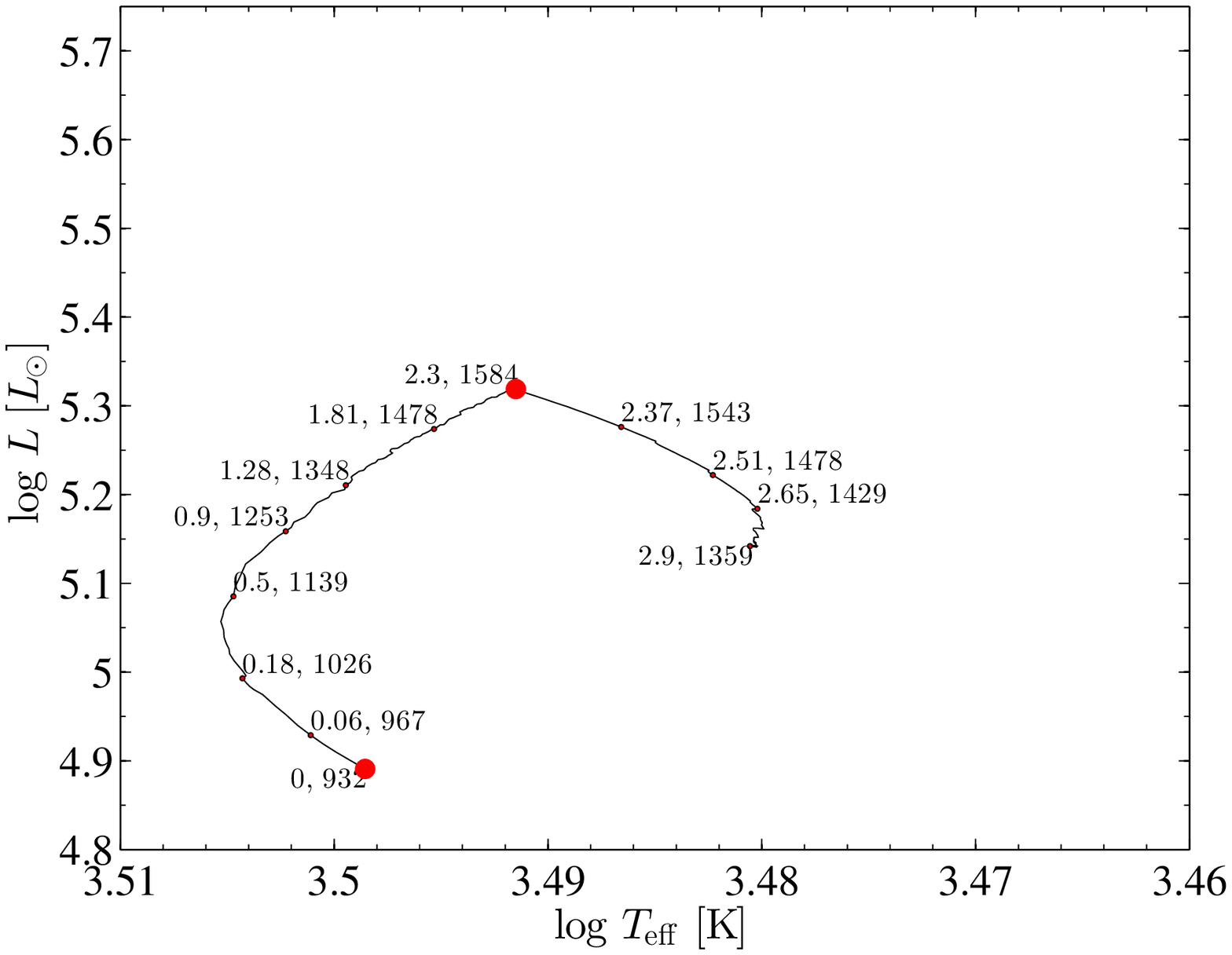}
       \hskip 0.3cm
   \includegraphics[width=0.65\textwidth]{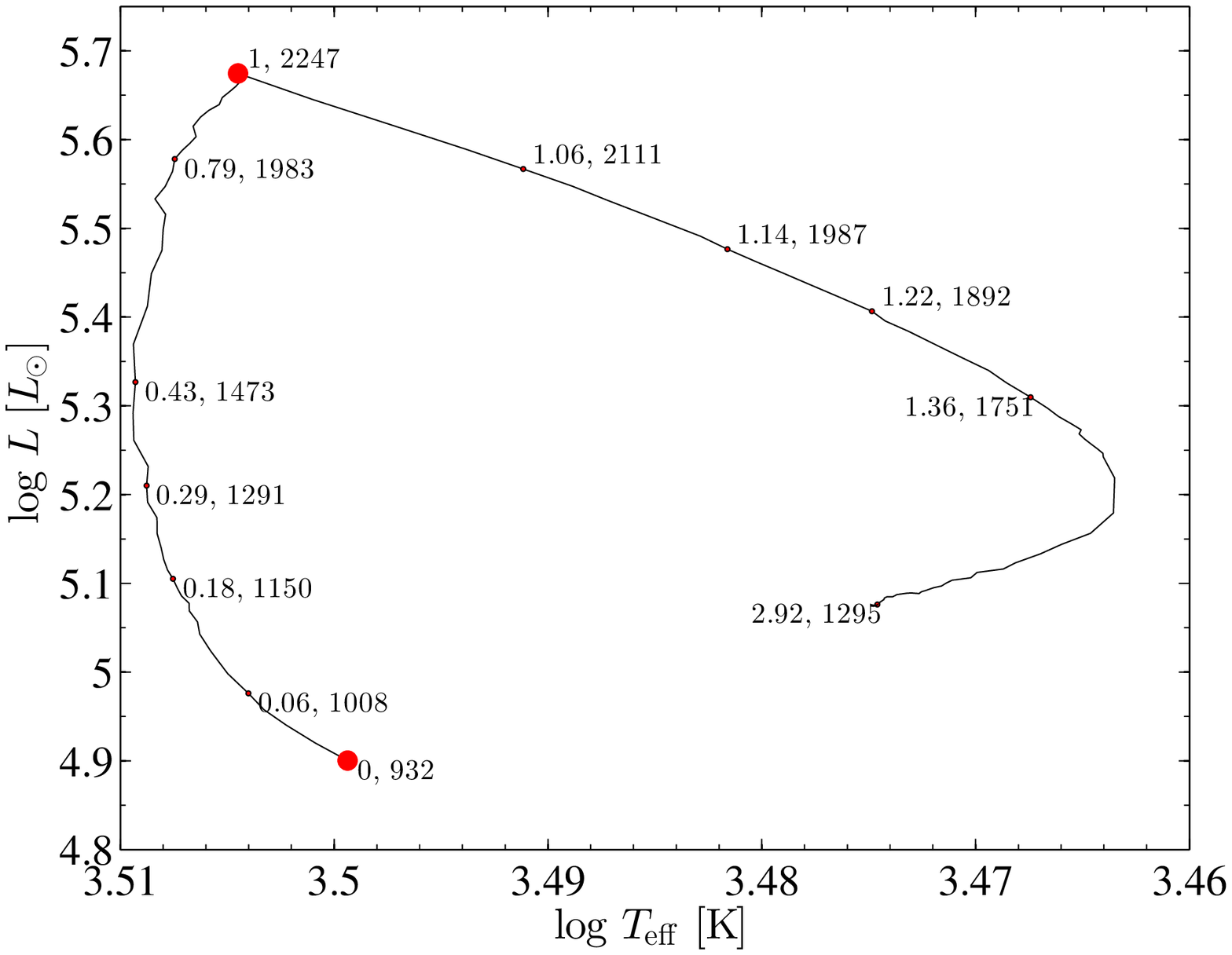}
\caption{The evolution of the $M_0=15M_\odot$ RSG model on the HR diagram during and after energy injection.
The lower red dot marks the beginning of wave-energy injection, and the upper red dot marks the end of the energy injection phase.
The numbers near several points along the evolution give
the time in years since energy injection started and the stellar radius. More details of the model are in Fig. \ref{figure:RSGModel}.
Upper panel is for $L_{\rm wave,n}=3.2 \times 10^5 L_\odot$, and the lower panel is for $L_{\rm wave,n}=1 \times 10^6 L_\odot$}
\label{figure:HR15}
\end{figure}

\begin{figure}[ht]
\centering
   \includegraphics[width=0.45\textwidth]{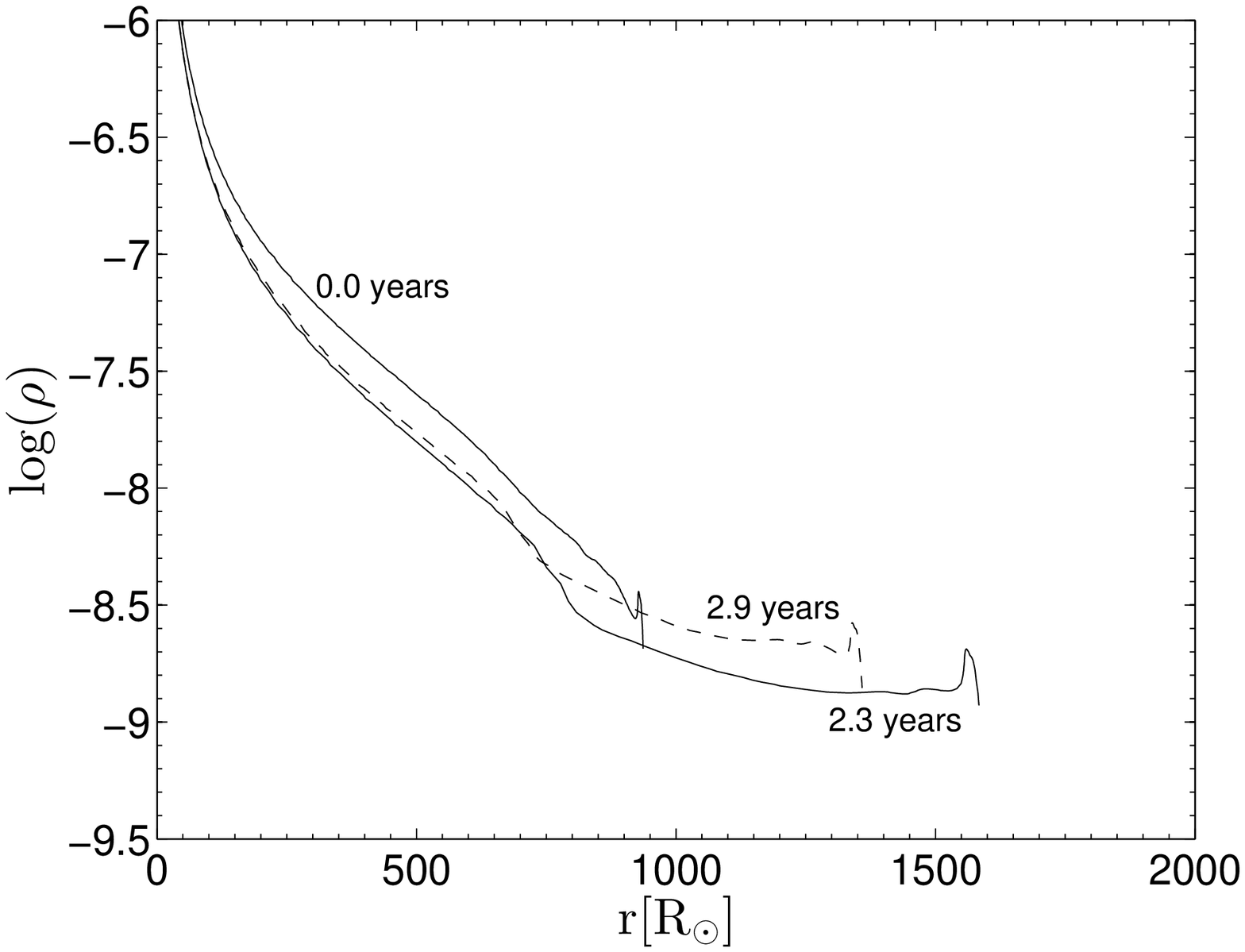}
       \hskip 0.3cm
   \includegraphics[width=0.45\textwidth]{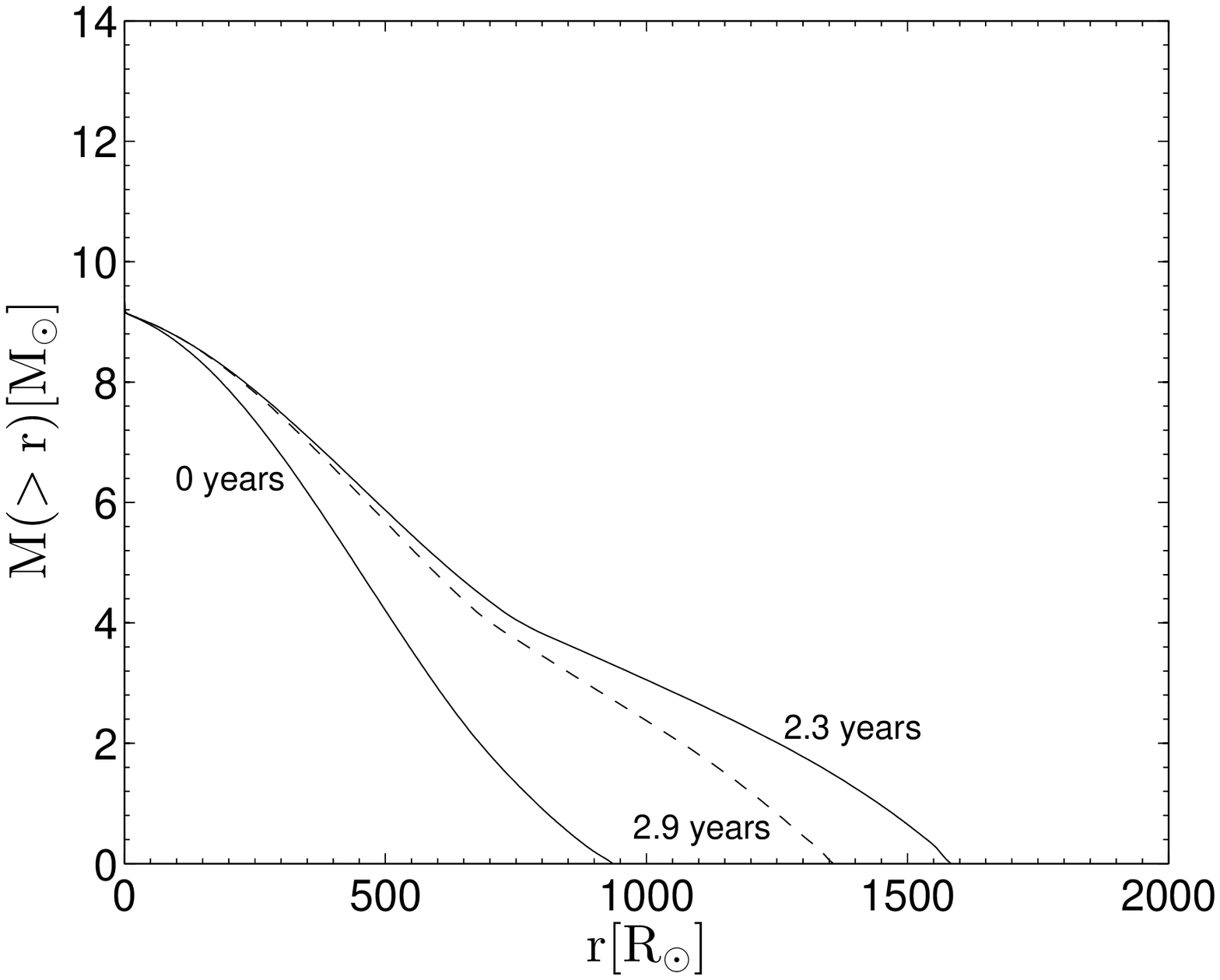}
\caption{Left: The stellar density profile at the beginning, $t=0$ and at the end, $t=2.3 M_\odot$, of energy injection phase,
and at $t=2.9 \yr$, for the $M_0=15M_\odot$ model with $L_{\rm wave,n}=3.2 \times 10^5 L_\odot$.
Note that the star shrinks after energy injection ceases.
Right: The mass outside radius $r$.
}
\label{figure:Density15}
\end{figure}

The RSG model is described in section \ref{sec:stellar} and is shown in Fig. \ref{figure:RSGModel} just before energy injection starts.
At core oxygen burning we take values for the non-rotating model of \cite{ShiodeQuataert2014}
$L_{\rm wave,n}=3.2 \times 10^5 L_\odot$, $t_{\rm wave,n}=2.3 \yr$, and $E_{\rm wave,n}=8.9 \times 10^{46} \erg$.
The evolution of the star in the HR diagram is presented in the upper panel of Fig. \ref{figure:HR15}, where we mark the time since the beginning of
energy injection and the stellar radius.
The density profiles and mass distributions at three times along the evolution of the $L_{\rm wave,n}=3.2 \times 10^5 L_\odot$
run are presented in Fig. \ref{figure:Density15}.

Let us first analyze the $L_{\rm wave,n}=3.2 \times 10^5 L_\odot$ run.
The star reaches a radius of $\sim 1600 R_\odot$ during the $2.3 \yr$ of the energy injection phase, and then contracts.
The free fall time for that radius is $\sim 1 \yr$, implying that the star can arrange itself for such a radius during the $2.3 \yr$ long energy injection phase.
As well, convective cells moving at the sound speed can transfer energy outward with the expanding envelope.
During the injection phase the star radiates a total energy of $E_{\rm rad}=4.4 \times 10^{46} \erg$ which is $\Delta E_{\rm rad}=2.2 \times 10^{46} \erg$
above what it would have radiated without energy injection.
Namely, the star radiates away $\Delta E_{\rm rad}/E_{\rm wave,n}=25 \%$ of the energy injected, and maintains $75 \%$ of it.
The star does not reach its Eddington luminosity of $4.5 \times 10^5  L_\odot$ (for electron scattering opacity).   

The initial energy injection radius is $r_{d}(0)=839 R_\odot$, and the mass above this radius is $M[>r_{d}(0)]= 0.58 M_\odot$.
If all injected energy is given to this mass, its initial velocity would be $124 \km \s^{-1}$, higher than the initial escape velocity from the star of
$74 \km \s^{-1}$.
The gravitational binding energy (not including thermal energy) of the mass $M[>r_{d}(0)]$ is $ 3.3 \times 10^{46}  \erg$, compared with the total injected energy of
$E_{\rm wave,n}=8.9 \times 10^{46} \erg$.
However, the injected energy does not go to eject the mass residing above $r_{d}(0)$.
As can be seen in Fig. \ref{figure:Density15} a huge envelope is inflated.
At the end of the injection phase the driven radius is $r_{d}(e) = 737 R_\odot$, and the mass above this radius is $\sim 4 M_\odot$, compared with $0.58 M_\odot$
above $r_{d}(0)$ when energy injection started.
The density in the entire envelope is reduced, and mass shells move outward to create an extended massive envelope. In reality the envelope inflation will
start earlier and will be more efficient due to the pressure of the waves (section \ref{sec:pwaveP}).

We conclude that for the above parameters the star can arrange itself to absorb all injected energy by inflating a huge massive envelope.
Namely, most of the injected energy is channelled to inflate the envelope.
The implication of such an extended envelope on a binary companion will be discussed in section \ref{sec:summary}.
It is true that in the way we use MESA dynamical effects are not included. But we set the goal here to show that the star can build itself into
a new structure accommodating the injected energy.
After injection stops, the star loses energy in radiation and starts to contract.

\begin{figure}[ht]
\centering
   \includegraphics[width=0.45\textwidth]{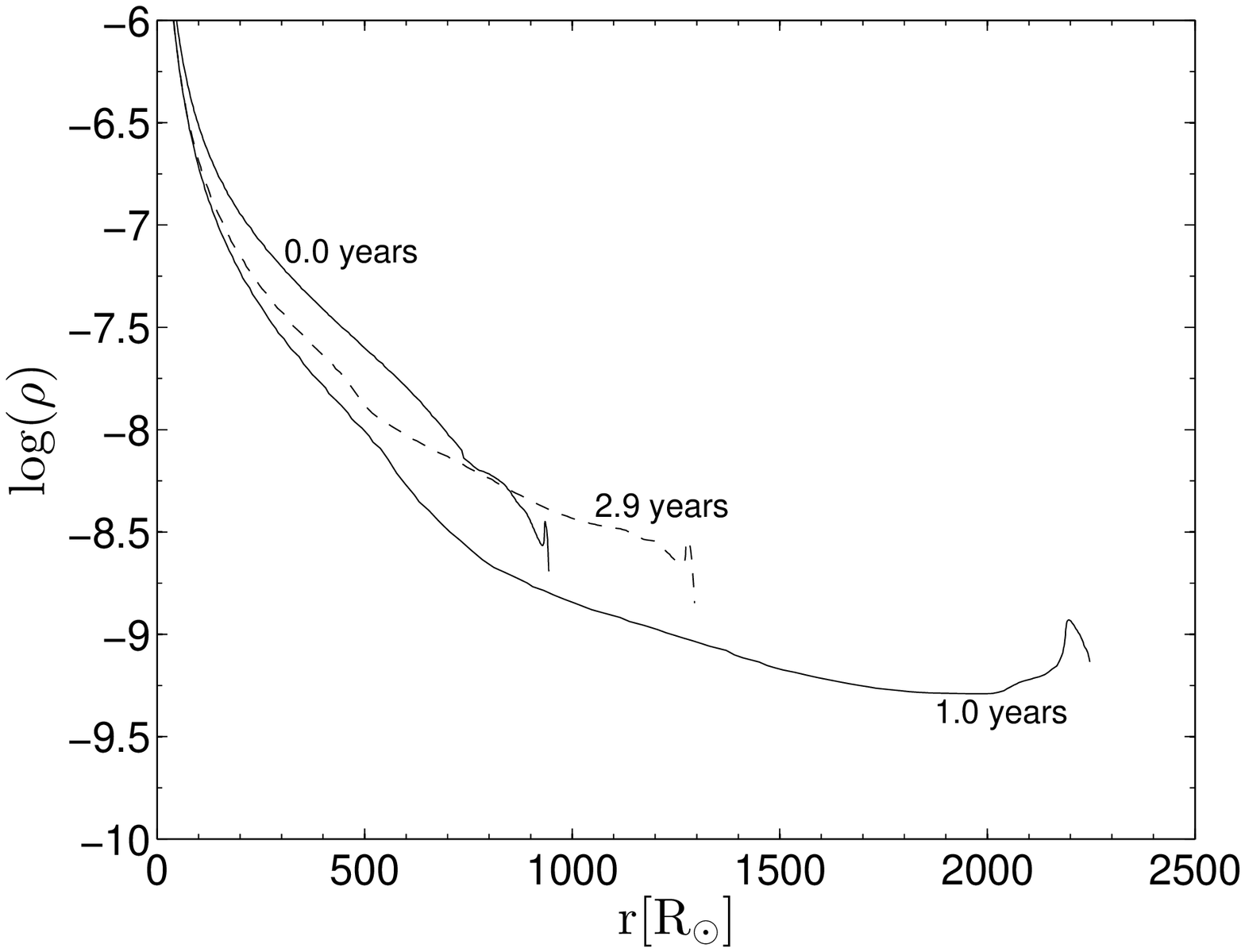}
       \hskip 0.3cm
   \includegraphics[width=0.45\textwidth]{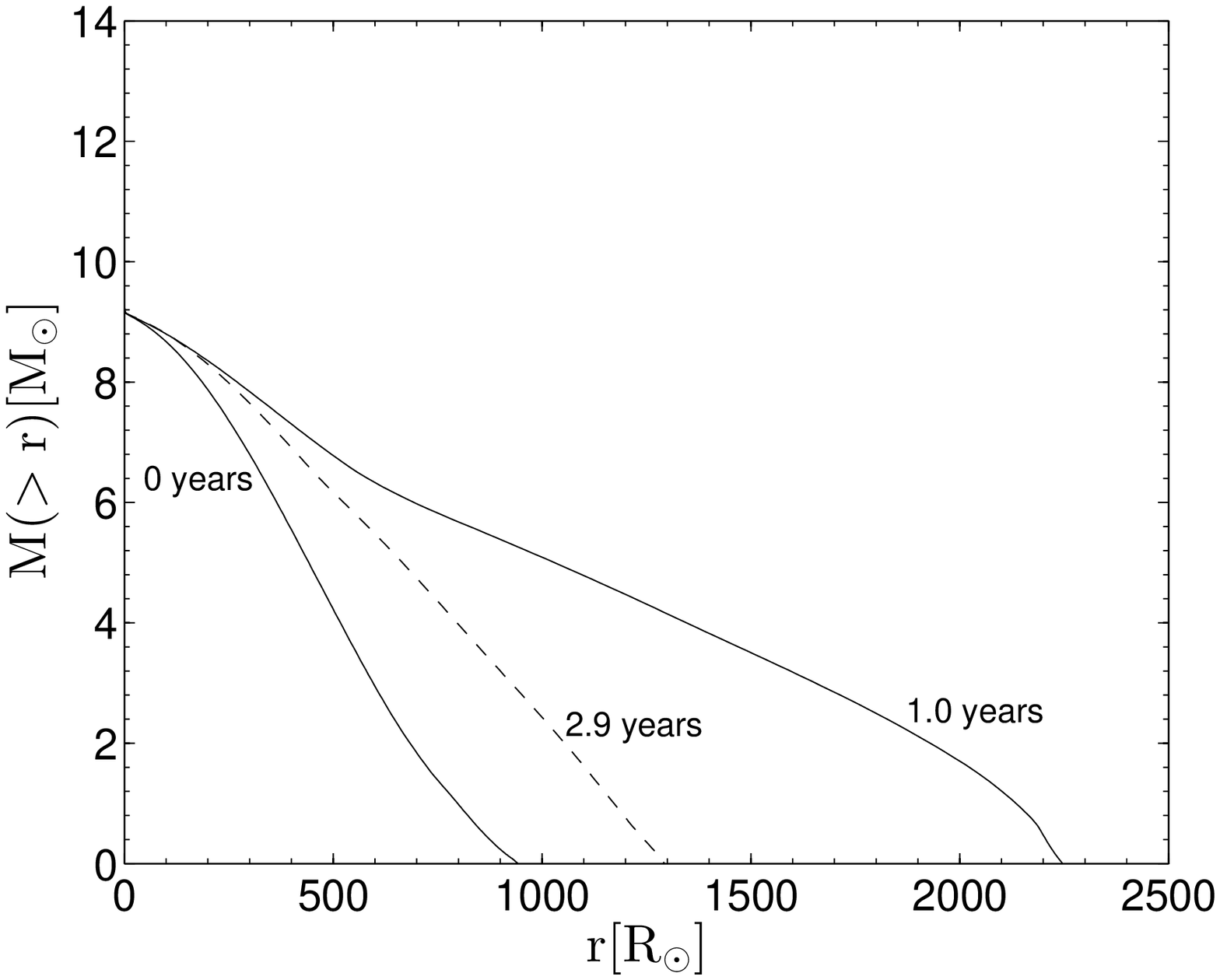}
\caption{Left: The stellar density profile at the beginning, $t=0$ and at the end, $t=1.0 \yr$, of energy injection phase,
and at $t=2.9 \yr$, for the $M_0=15M_\odot$ model with $L_{\rm wave,n}=1 \times 10^6 L_\odot$.
Right: The mass outside radius $r$.
}
\label{figure:Density15high}
\end{figure}

If energy injection power is sufficiently high the star has a problem to maintain a stable structure.
We demonstrate this with a wave power of $L_{\rm wave,n}=1 \times 10^6 L_\odot$, that is three times higher than given by of \cite{ShiodeQuataert2014}.
We inject the energy during one year.
The evolution on the HR diagram is shown in the lower panel of Fig. \ref{figure:HR15}, and the density profile and mass distribution at three times are shown
in Fig. \ref{figure:Density15high}.
The initial energy injection radius in this case is $r_{d}(0)= 780 R_\odot$, and the mass above this radius is $M[>r_{d}(0)]= 1.06 M_\odot$.
At the end of the injection phase ($t=1 \yr$) the energy injection radius is $r_{d}(e) = 520 R_\odot$, and the mass above this radius is $\sim 6.9 M_\odot$.

The calculation of the stellar structure in this case of $L_{\rm wave,n}=1 \times 10^6 L_\odot$ is not fully consistent
as the dynamical time for radii $>1600 R_\odot$  is larger than the injection period of $1 \yr$.
Even if some mass is ejected with high velocities, we don't expect the mass to reach the escape speed.
For example, the mass residing above $1600 R_\odot$ at the end of the injection phase is $\sim 3.2  M_\odot$.
With the entire injected energy the velocity of this gas would be $62 \km \s^{-1}$, which is less than the escape velocity from the star at $t=0$.
Even in this extreme case we don't expect major mass ejection, but rather huge massive envelope inflation on top of which there is a
zone where parcels of dense gas rise and fall back. Such an `effervescent zone' allows for a strong interaction with a binary companion \citep{Soker2008},
to be discussed in section \ref{sec:summary}.
We note that the stellar surface temperature is low, and hence opacity is much lower than electron scattering opacity,
and the star is below its Eddington luminosity, although very close to it.

{{{
The inflated envelopes obtained here are similar in some respects to the inflated envelope around WR stars \citep{Petrovicetal2006, Grfeneretal2012}.
\cite{Petrovicetal2006} and \cite{Grfeneretal2012} show that an inflated envelope, up to $\sim 2-4$ times the original stellar radius, can be developed around WR stars
when the stars are close to their Eddington luminosity and the mass loss rate is low.
The outer region of the inflated envelope is convective and possesses a large density inversion region, much as seen here in the left panels of Figs.
\ref{figure:Density15} and \ref{figure:Density15high} at maximum envelope inflation.
The situation here, we argue, is similar in that the star is close to its Eddington luminosity limit, and mass loss rate is small.
While \cite{Petrovicetal2006} assume low mass loss rate, here we argue there is no time for regular mass loss process to develop.
The envelope then reacts by expanding.
}}}

\subsection{A $40 M_\odot$ blue supergiant}
\label{subsec:BSG}

We turn to a blue supergiant (BSG) model with an initial mass of $M_0=40 M_\odot$, and a final mass of $19.7 M_\odot$. The stellar radius when injection starts is $191 R_\odot$
(section \ref{sec:stellar}).
Our model has a large initial radius, but the mass between $50 R_\odot$ and the stellar surface is only
$0.33 M_\odot$. This low mass extended envelope does not affect our results much.
At core oxygen burning values from table 2 of \cite{ShiodeQuataert2014} are
$L_{\rm wave,n}=4 \times 10^7 L_\odot$, $t_{\rm wave,n}=0.092 \yr$, and $E_{\rm wave,n}=4.5 \times 10^{47} \erg$.
For numerical reason we start wave injection with $L_{\rm wave,n}=4 \times 10^7 L_\odot$
just before core-oxygen burning, and continue till core collapse $t_{\rm wave,n}=0.089 \yr$ later; the total
wave energy injected is $E_{\rm wave,n}=4.3 \times 10^{47} \erg$.
During the wave energy injection phase the star has radiated an extra (above its regular radiation) energy of
$8.4 \times 10^{46} \erg$, which is $20 \%$ of the injected wave energy.

\begin{figure}[ht]
\centering
\includegraphics[width=0.65\textwidth]{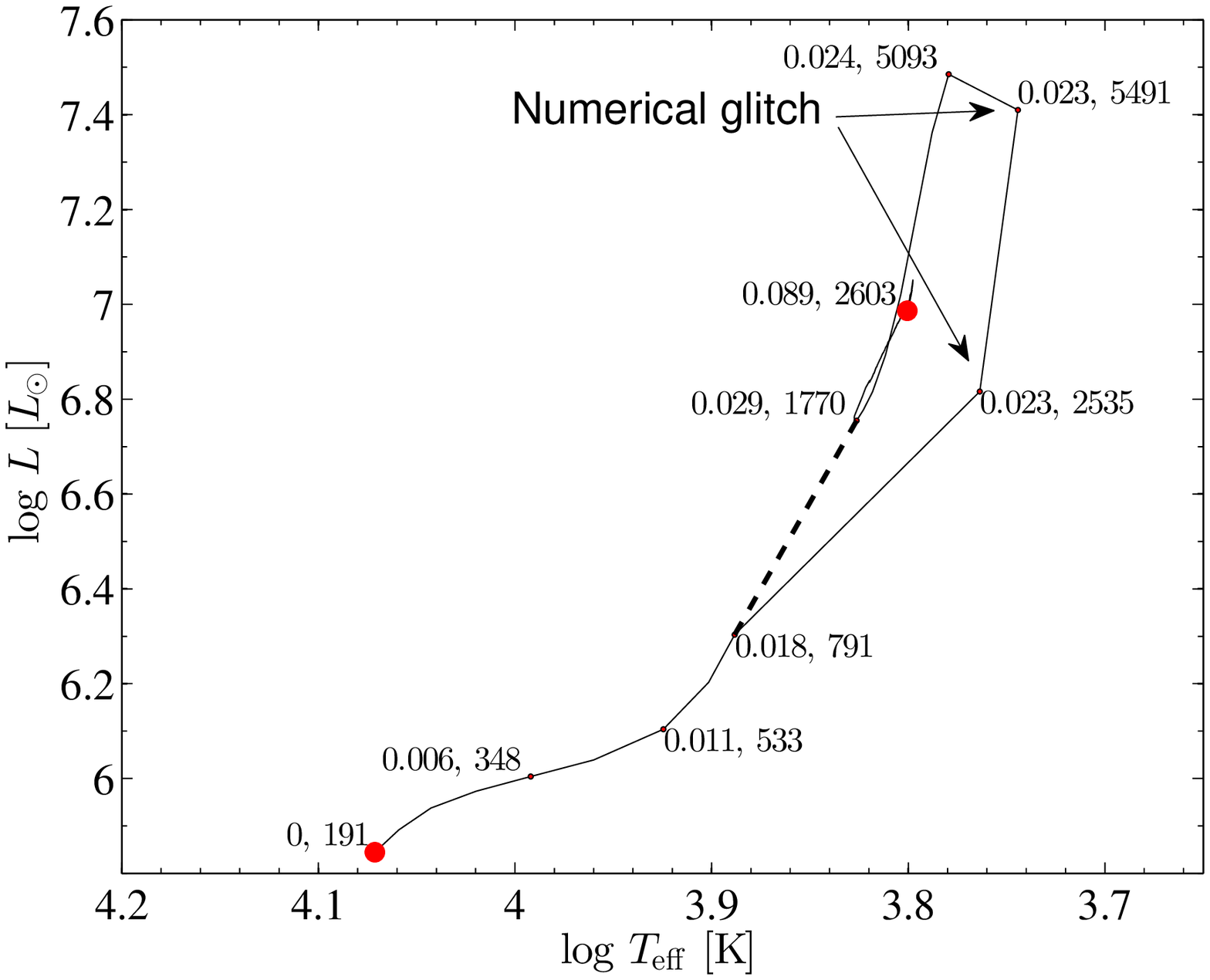}
       \hskip 0.3cm
\caption{The evolution of the $M_0=40M_\odot$ model on the HR diagram during wave energy deposition. The lower red dot marks the beginning of wave-energy injection,
and the upper red dot marks the end of the energy injection phase, when the core starts to collapse.
The numbers near several points along the evolution give the time in years since beginning of energy injection and the stellar radius.
The evolution from $0.023 \yr$ to $0.029 \yr$ is a numerical effect (`glitch') during which the code arranges itself, and we take the evolution to be more
or less along the thick dashed line.  }
\label{figure:HR40}
\end{figure}

\begin{figure}[ht]
\centering
\includegraphics[width=0.45\textwidth]{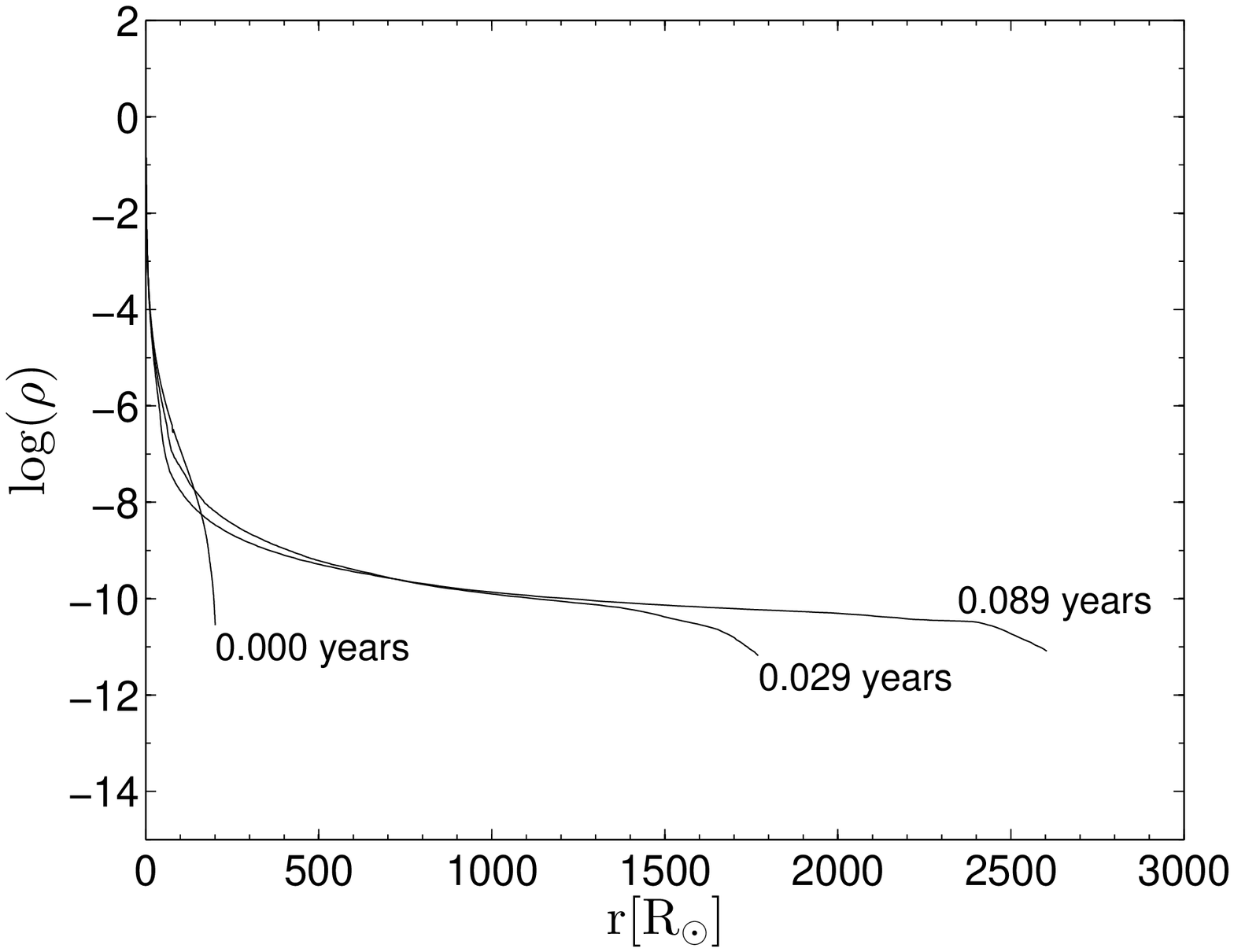}
       \hskip 0.3cm
\includegraphics[width=0.45\textwidth]{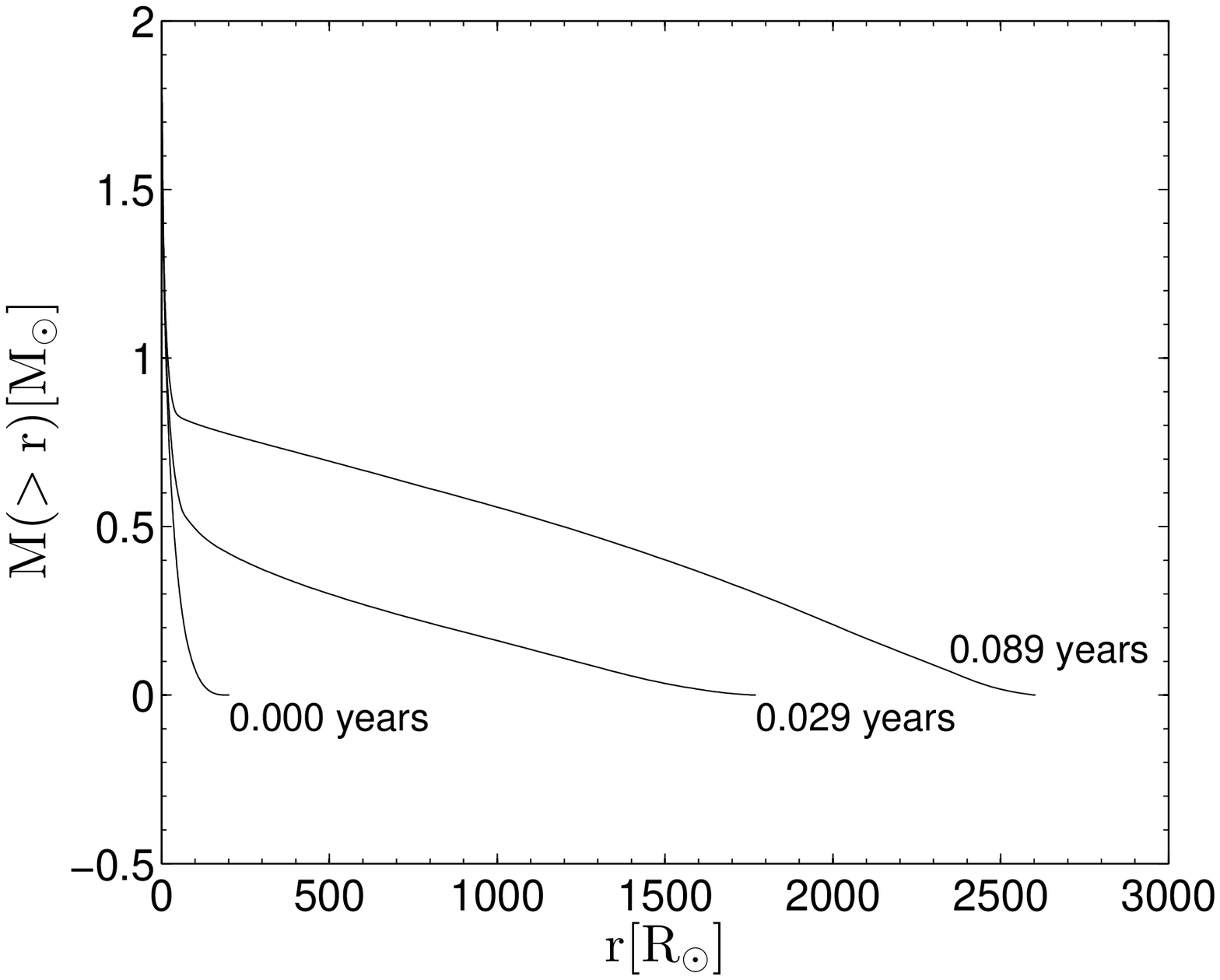}
\caption{Left: The stellar density profile at the beginning, $t=0$, $t=0.029 \yr$,  and at the end, $t=0.089 \yr$, of energy injection phase,
for the $M_0=40M_\odot$ model with $L_{\rm wave,n}=4 \times 10^7 L_\odot$.
Right: The mass outside radius $r$.
}
\label{figure:Density40}
\end{figure}

The evolution of the star in the HR diagram is presented in Fig. \ref{figure:HR40},
where we mark the time in years since the beginning of energy injection and the stellar radius.
We take the rapid evolution from $t=0.023 \yr$ to $t=0.029 \yr$ to be a numerical effect where the code arranges itself, and ignore this part.
We take the evolution to occur along the thick dashed line.
As the effective temperature falls below $10^4 \K$, opacity drops below that of electron scattering and its value at the end of the energy injection phase is
$\kappa=0.02 \cm^2 \g^{-1}$.
This implies that the star does not exceed its Eddington luminosity, but it is very close to it.
The density profiles and mass distribution at three times along the evolution are presented in Fig. \ref{figure:Density40}.

The star expends very rapidly, on a time scale shorter than its free fall time on the initial surface, $0.03 \yr$.
This is not fully consistent as the version of MESA we use does not have a dynamical part and cannot follow mass ejection.
In the region where energy is injected the dynamical time is shorter than the injection period, and this region and inward is treated consistently.
A large part of the injected energy is absorbed by stellar mass at radii of $r< 100 R_\odot$
(right panel of Fig. \ref{figure:Density40}), where the dynamical time is $<0.013 \yr$.
The outer regions are more likely to expand slowly, with velocity less than the escape velocity, than maintain hydrostatic equilibrium.
We can safely conclude that the results show that the expanding envelope can absorb a large fraction of the injected energy.
Namely, the outcome of wave energy deposition is not a fast and light mass ejection, but rather a slowly expanding massive envelope.
This holds both for our RSG model studied in section \ref{subsec:RSG} and for the BSG studied here.

Let us quantify the above statement.
The initial energy injection radius is $r_{d}(0)=80 R_\odot$, and the mass above this radius is
$M[>r_{d}(0)]= 0.13 M_\odot$.
The inner boundary of the energy injection zone at the end, $t= 0.089 \yr$, is $41 R_\odot$,
and the mass above this radius is $M[>r_{d}(0.089)]= 0.84 M_\odot$.
If the entire injected energy of $4.3 \times 10^{47} \erg$ is chandelled to kinetic energy of this mass,
its velocity would be $ 227 \km \s^{-1}$. This is the escape velocity from a radius of $140 R_\odot$.
Since the energy is deposited while the mass is at smaller radii, the mass will not escape.
The mass might form an extended `effervescent zone'. An explosion will occur before the outer region manages
to fall back.

As most of the energy is channelled to envelope expansion, no bright outburst with massive mass ejection will take place from this process.
Our results give a bright star, but we expect that in a dynamical calculation there will be no time to transfer energy, and the luminosity will be much lower.
The extra energy that is not radiated will be in the kinetic energy of the slowly expanding envelope.
A way to form an outburst is if a massive main sequence companion enter (collide with) the extended envelope.
A strong shock is formed, and in addition the companion can accrete mass and liberate more gravitational energy.
The companion needs to be on an orbital separation of $a \simeq 3 AU$, preferentially on an eccentric orbit.
Finally, we note that the progenitor of SN~1987A was a BSG, although with a lower initial mass of $\sim 20 M_\odot$,
but no pre-explosion outburst was observed from it. So definitely not all CCSNe have such a phase of large expansion.

\section{DISCUSSION AND SUMMARY}
\label{sec:summary}

The heterogenous group of erupting objects having luminosity between novae and supernovae (SNe) is composed mainly of intermediate luminosity optical transients (ILOTs; also termed intermediate
luminosity red transients or red luminous novae) and major outbursts of luminous blue variables (LBV).
Now there are evidences for pre-explosion outbursts (PEOs) that reside in this regime as well.
There are two classes of models to account for ILOTs, LBV major outbursts, and PEOs.
One school attributes the eruptions and their energy source to single star evolution, e.g., the porous-atmosphere
model that was suggested by \cite{Shaviv2000} for the super-Eddington nineteenth century Great Eruption of $\eta$ Car,
the single star model of SN~2008S and NGC~300~OT \citep{Kochanek2011}, and the model of \cite{ShiodeQuataert2014} to the PEO of SN~2010mc.
\cite{Soker2004} already argued that single star models, such as the rotation-based singe star scenario of \cite{Smithetal2003}, fail in explaining the bipolar nebula of
$\eta$ Car.
\cite{McleySoker2014} argue that single-star models for ILOTs of evolved giant stars, such as NGC~300~OT and SN~2008S,  encounter severe difficulties.

The other school argues that the source of energy is gravitational energy released by either mass transfer in a binary system or by stellar merger.
Although the event might be triggered by an instability in one star, binary interaction is at the heart of these eruptions.
Examples include mass transfer processes as in the binary model for the Great Eruption of $\eta$ Car
\citep{Soker2001, KashiSoker2010}, NGC~300~OT \citep{Kashietal2010}, and in the early outbursts of SN~2009ip \citep{SokerKashi2013}.
A binary merger process occurred for example in V1309 Scorpii \citep{Tylendetal2011}.

A strong binary interaction, including violent mass transfer to a binary companion, can take place when the giant star suffers rapid expansion and inflates a massive extended envelope.
The envelope of LBV stars has radiative and convective layers, such that magnetic energy can be stored in the radiative zone \citep{HarpazSoker2009}.
The release of the magnetic energy can lead to envelope distortion that leads to huge envelope expansion and mass loss \citep{HarpazSoker2009} that facilitate binary interaction,
such as in $\eta$ Car.

In the present paper we looked at a different triggering of envelope expansion.
We studied the process suggested by \cite{QuataertShiode2012} and \cite{ShiodeQuataert2014},
where waves excited by the vigorous burning in the core of pre-exploding supernovae years to hours before explosion, dissipates in the outer envelope.
\cite{Ofeketal2013b} also preferred a single star model for the PEO of SN~2010mc and adopted the porous-atmosphere model \citep{Shaviv2000, Shaviv2001}.
We instead argued in this paper that the powerful waves lead to envelope expansion rather than mass ejection. To account for the PEO, a binary companion that interacts
with the extended envelope powers the PEO \citep{Soker2013}.
The high binary fraction among massive stars, e.g. \cite{Kobulnickyetal2014}, makes binary interaction of a hugely expanding giant star very likely.

In section \ref{sec:pwaveP} we estimated the energy and pressure content of the waves, and presented arguments that
show that most of the energy carried by the waves will lead to envelope expansion rather than mass ejection at the escape speed.
We showed that in red supergiants (RSG) the waves start to increase the pressure, and hence
cause envelope expansion deeper in the envelope than where convection driven by waves become supersonic ($r_d$).
The envelope will expand, rather than ejects its outer part.
In section \ref{subsec:Meject} we strengthen this conclusion by simple arguments regarding the dissipation of the wave energy.

In section \ref{sec:numerical} we used the MESA code to study energy injection over about a year for a model of a RSG, and over about a month into a blue supergiant (BSG) stellar model.
We used several simplifying assumptions as to the place of energy injections.
We found that for the wave powers as listed by \cite{ShiodeQuataert2014} the stars can absorb the energy and arrange themselves in a structure with a very extended massive envelope.
This extended envelope is likely to have a slow expansion, and the SN explosion might occur before the outer extended envelope sets into a full hydrostatic equilibrium.
It is with such a very extended envelope that a binary companion can interact. Due to the large mass at large radii, if the companion orbits through the envelope it can accrete a large
fraction of the mass in the extended envelope.

\cite{Moriyaetal2014} modelled the light curve of 11 Type IIn supernova and derived the circumstellar medium properties.
In some cases the high mass loss rate from the progenitor started at least 60 years prior to explosion.
If caused by nuclear burning, it requires the stage of carbon burning. The carbon burning has insufficient energy to cause
mass loss rate by wave dissipation, but it can inflate the envelope.
Binary interaction, e.g., by tidal interaction, can take place and enhance mass loss rate.
A low mass companion might spiral-in due to tidal interaction. It spins-up the envelope and increases the mass loss rate.
As mass loss proceeds the envelope slows down and mass loss decreases.
Such a scenario might account for the decreases in mass loss rate found for one of the systems studied by \cite{Moriyaetal2014}.
We predict that the common morphology of the circumstellar matter of SN IIn is bipolar.

{{{ We thank the referee, Takashi Moriya, for helpful comments. }}}
This research was supported by the Asher Fund for Space Research at the Technion, the US-Israel Binational Science Foundation,
and a generous grant from the president of the Technion Prof. Peretz Lavie.


\begin{thebibliography}

\bibitem[Bear et al.(2011)]{Bearetal2011} Bear, E., Soker, N., \& Harpaz, A.\ 2011, \apjl, 733, L44

\bibitem[Boothroyd \& Sackmann(1988)]{Boothroyd1988} Boothroyd, A.~I., \& Sackmann, I.-J.\ 1988, \apj, 328, 671

\bibitem[Campbell \& Lattanzio(2008)]{Campbell2008} Campbell, S.~W., \& Lattanzio, J.~C.\ 2008, First Stars III, 990, 315

\bibitem[Campbell et al.(2010)]{Campbell2010} Campbell, S.~W., Lugaro, M., \&  Karakas, A.~I.\ 2010, \aap, 522, L6

{{{ \bibitem[Glebbeek et al.(2009)]{Glebbeeketal2009} Glebbeek, E., Gaburov, E., de Mink, S.~E., Pols, O.~R., \& Portegies Zwart, S.~F.\ 2009, \aap, 497, 255  }}}

{{{ \bibitem[Gr{\"a}fener et al.(2012)]{Grfeneretal2012} Gr{\"a}fener, G., Owocki, S.~P., \& Vink, J.~S.\ 2012, \aap, 538, A40  }}}

\bibitem[Harpaz \& Soker(2009)]{HarpazSoker2009} Harpaz, A., \& Soker, N.\ 2009, \na, 14, 539

\bibitem[Kashi et al.(2010)]{Kashietal2010} Kashi, A., Frankowski, A., \& Soker, N.\ 2010, \apjl, 709, L11

\bibitem[Kashi \& Soker(2010)]{KashiSoker2010} Kashi, A., \& Soker, N.\ 2010, \apj, 723, 602

\bibitem[Kobulnicky et al.(2014)]{Kobulnickyetal2014} Kobulnicky, H.~A., Kiminki, D.~C., Lundquist, M.~J., et al.\ 2014, arXiv:1406.6655

\bibitem[Kochanek(2011)]{Kochanek2011} Kochanek, C.~S.\ 2011, \apj, 741, 37

\bibitem[Mauron \& Josselin(2011)]{MauronJosselin2011} Mauron, N., \& Josselin, E.\ 2011, \aap, 526, A156

\bibitem[Mcley \& Soker(2014)]{McleySoker2014} Mcley, L., \& Soker, N.\ 2014, \mnras, 440, 582

\bibitem[Mocak et al.(2008)]{Mocak2008} Moc{\'a}k, M., M{\"u}ller, E., Weiss, A., \& Kifonidis, K.\ 2008, \aap, 490, 265

\bibitem[Mocak (2009)]{Mocak2009} Moc{\'a}k, M.\ 2009, Ph.D.~Thesis.

\bibitem[Mocak et al.(2010)]{Mocak2010} Moc{\'a}k, M., Campbell, S.~W., M{\"u}ller, E., \& Kifonidis, K.\ 2010, \aap, 520, A114

\bibitem[Moriya(2014)]{Moriya2014} Moriya, T.~J.\ 2014, arXiv:1403.2731

\bibitem[Moriya et al.(2014)]{Moriyaetal2014} Moriya, T.~J., Maeda, K., Taddia, F., Sollerman, J., Blinnikov, S. I., Sorokina, E. I.\ 2014, arXiv:1401.4893

{{{ \bibitem[Nugis \& Lamers(2000)]{Nugis2000} {Nugis}, T. \& {Lamers}, H.~J.~G.~L.~M.\ 2000, \aap, 360, 227 }}}

\bibitem[Ofek et al.(2013)]{Ofeketal2013b} Ofek, E. O. et al.\ 2013, \nat 494, 65

\bibitem[Owocki et al.(2004)]{Owockietal2004} Owocki, S.~P., Gayley, K.~G., \& Shaviv, N.~J.\ 2004, \apj, 616, 525

\bibitem[Paxton et al.(2011)]{Paxton2011} Paxton, B., Bildsten, L., Dotter, A., et al.\ 2011, \apjs, 192, 3

{{{  \bibitem[Petrovic et al.(2006)]{Petrovicetal2006} Petrovic, J., Pols, O., \& Langer, N.\ 2006, \aap, 450, 219   }}}

\bibitem[Quataert \& Shiode(2012)]{QuataertShiode2012} Quataert, E., \& Shiode, J.\ 2012, \mnras, 423, L92

\bibitem[Schlattl et al.(2001)]{Schlattl2001} Schlattl, H., Cassisi, S., Salaris, M., \& Weiss, A.\ 2001, \apj, 559, 1082

\bibitem[Shaviv(2000)]{Shaviv2000} Shaviv, N.~J.\ 2000, \apjl, 532, L137

\bibitem[Shaviv(2001)]{Shaviv2001} Shaviv, N.~J.\ 2001, \mnras, 326, 126

\bibitem[Shiode \& Quataert(2014)]{ShiodeQuataert2014} Shiode, J.~H., \& Quataert, E.\ 2014, \apj, 780, 96

\bibitem[Smith et al.(2003)]{Smithetal2003} Smith, N., Davidson, K., Gull, T.~R., Ishibashi, K., \& Hillier, D.~J.\ 2003, \apj, 586, 432

\bibitem[Soker(1992)]{Soker1992} Soker, N.\ 1992, \apj, 386, 190

\bibitem[Soker(1993)]{Soker1993} Soker, N.\ 1993, \apj, 417, 347

\bibitem[Soker(2001)]{Soker2001} Soker, N.\ 2001, \mnras, 325, 584

\bibitem[Soker(2004)]{Soker2004} Soker, N.\ 2004, \apj, 612, 1060

\bibitem[Soker(2008)]{Soker2008} Soker, N.\ 2008, \na, 13, 491

\bibitem[Soker(2013)]{Soker2013} Soker, N.\ 2013, arXiv:1302.5037

\bibitem[Soker \& Kashi(2013)]{SokerKashi2013} Soker, N., \& Kashi, A.\ 2013, \apjl, 764, L6

\bibitem[Suda \& Fujimoto(2010)]{Suda2010} Suda, T., \& Fujimoto, M.~Y.\ 2010, \mnras, 405, 177

\bibitem[Tylenda et al.(2011)]{Tylendetal2011} Tylenda, R., Hajduk, M., Kami{\'n}ski, T.,  et al.\ 2011, \aap, 528, 114

{{{ \bibitem[Vink et al.(2001)]{Vink2001} {Vink}, J.~S., {de Koter}, A., \& {Lamers}, H.~J.~G.~L.~M.\ 2001, \aap, 369, 574 }}}

\end{thebibliography}
\end{document}